\documentclass[10pt]{iopart}		
\usepackage{graphicx}

\newcommand{\rf}     [1] {~\cite{#1}}
\newcommand{\FPoper}{Perron-Frobenius oper\-ator} 
\newcommand{\Lop}{{\cal L}}        
\newcommand{\dzeta}{dyn\-am\-ic\-al zeta func\-tion}
\newcommand{\fd}{spec\-tral det\-er\-min\-ant}
\newcommand{\Lnoise}[1]{{\cal L}_\sigma^{#1}}   
\newcommand{\beq}{\begin{equation}}
\newcommand{\refref} [1] {ref.~\cite{#1}}
\newcommand{\reffig} [1] {fig.~\ref{#1}}
\newcommand{\eeq}{\end{equation}}
\newcommand{\ee}[1] {\label{#1} \end{equation}}
\newcommand{\pde}{\partial}
\newcommand{\bea}{\begin{eqnarray}}
\newcommand{\continue}{\nonumber \\ }
\newcommand{\eea}{\end{eqnarray}}
\newcommand{\cl}[1]{{n_{#1}}}   
\newcommand{\ceq}{\nonumber \\ & & }
\newcommand{\refeq}  [1] {(\ref{#1})}
\newcommand{\inFix}[1]{{\in \mbox{\footnotesize Fix}f^{#1}}}
\newcommand{\averx}{\int dx\,}
\newcommand{\nnu}{\nonumber}

\newcommand{\refsect}[1] {sect.~\ref{#1}}

\begin{document}
\renewcommand{\baselinestretch}{1.2}
\jl{8}  
\title[Stochastic trace formulas]{Trace formulas for stochastic
evolution operators: smooth conjugation method}

\author{Predrag Cvitanovi\'c\dag,
	C P Dettmann\dag,
	Ronnie Mainieri\ddag\
	and
	G\'abor Vattay\S
}
\address{\dag\ Center for Chaos and Turbulence Studies,
	Niels Bohr Institute\\
	Blegdamsvej~17, DK-2100 Copenhagen \O\ }
\address{\ddag\
Theoretical Division, MS B213\\  
Los Alamos National Laboratory,  
Los Alamos, NM 87545 }
\address{\S\
Dept. Solid State Physics\\
E\"otv\"os University,
Muzeum krt. 6-8,
H-1088 Budapest   } 

\date{\today}

\begin{abstract}
The trace formula for the evolution operator
associated with nonlinear stochastic flows
with weak additive noise is cast in the path integral formalism.
We integrate over the neighborhood of a given
saddlepoint exactly by means of a smooth conjugacy,
a locally analytic nonlinear change of
field variables. The perturbative corrections
are transfered to the corresponding Jacobian, which we
expand in terms of the conjugating function, rather than the 
action used in defining the path integral.
The new perturbative expansion  which follows by a recursive
evaluation of derivatives appears more compact than the standard
Feynman diagram perturbation theory.
The result is a stochastic analog of the Gutzwiller trace formula
with  the ``$\hbar$'' corrections computed an order higher than
what has so far been attainable in stochastic and quantum-mechanical
applications.
\end{abstract}

\noindent
\\
{\bf keywords:} 
noise, stochastic dynamics, 
Fokker-Planck operator, 
cycle expansions, 
periodic orbits,
semiclassical expansions, 
perturbative expansions, 
trace formulas,
spectral determinants, 
zeta functions.

\section{Introduction}
The study of stochastic perturbations of dynamical systems is important
in applications to realistic systems since their description contains
uncertainties due to degrees of freedom that are not included, either
because they are not measurable, or because their inclusion would
complicate the analysis.  
A small stochastic perturbation can
smooth the singular distributions inherent in many nonlinear dynamical
systems.  This makes it easier to study averages over such distributions as 
well defined deterministic limits of smooth stochastic distributions.
Besides being an ubiquitous fact of life - any dynamics in nature
is stochastic to some degree, however weak -- stochastic processes 
offer us a great freedom in picking problems and testing ideas. 

The properties of the weakly stochastic system are here
determined from the unstable periodic orbits of the unperturbed
(deterministic) system, decorated by the stochastic corrections.
The central object in the theory, the trace of the evolution operator,
is a discrete path integral, similar to those found in field theory
and statistical mechanics.  The weak noise perturbation theory, likewise, resembles
perturbative field theory, and in the preceding paper\rf{noisy_Fred}
we developed such a perturbation theory for trace formulas for weakly
stochastic chaotic dynamics in the standard field-theoretic language
of Feynman diagrams.

Here we approach the same problem from an althogether different
direction; the key idea of flattening the neighborhood of
a saddlepoint can be traced back to 
Poincar\'e\rf{poincare},
and is perhaps not something that a field theorist would instinctively
hark to as a method of computing perturbative corrections.
In the Feynman diagram approach\rf{noisy_Fred} we 
observed that the sums of diagrams simplify
for saddlepoints corresponding to repeats of shorter periodic orbits,
and were surprised by
the compactness of the order $\sigma^2$ correction.
Here we explain this simplification in geometric terms that might be
applicable to more general field theoretic problems.

\section{Stochastic evolution operator}

The periodic orbit theory allows us to calculate long time averages
in a chaotic system as expansions in terms of the periodic orbits
(cycles) of the system.  
The simplest example is provided by the {\FPoper} 
\[
\Lop \rho(x')=\int dx\,\delta(f(x)-x')\rho(x)
\]
for a {\em deterministic} map $f(x)$ which maps a density distribution
$\rho(x)$ forward in time.
The periodic orbit theory relates the spectrum of this operator and
its weighted evolution operator generalizations to the periodic orbits
via trace formulas, \dzeta s and \fd s\rf{PG97,QCcourse}. 
For quantum mechanics the periodic orbit theory is exact on the
semiclassical level\rf{gutbook}, whereas the quintessentially quantum
effects such as creeping, tunneling
and diffraction have to be included as corrections. In particular,
the higher order $\hbar$ corrections can be computed
perturbatively by means of Feynman diagrammatic expansions\rf{alonso1}.
Our
purpose here is to develop the parallel theory for {\em stochastic}
dynamics.
In case at hand, already a discrete time 1-dimensional 
discrete Langevin equation\rf{vk,LM94},
\begin{equation}
x_{n+1}=f(x_n)+\sigma\xi_n
\,,\label{Langevin}
\end{equation}
with $\xi_n$ independent normalized
random variables,
suffices to reveal the structure of the perturbative corrections.

We treat a chaotic system with 
weak external noise by 
replacing the deterministic evoluton $\delta$-function kernel 
by $\Lnoise{}$,  the Fokker-Planck
kernel corresponding to (\ref{Langevin}),
a sharply peaked noise distribution function 
\beq
\Lnoise{}(x',x) =\delta_\sigma(f(x)-x')
\,.
\ee{Lnoise}
In \refref{noisy_Fred} we have treated the problem
of computing the spectrum of this operator by standard
field-theoretic Feynman diagram expansions. This time we 
formulate the perturbative expansion in terms of
smooth conjugacies and
recursively evaluated derivatives. The procedure, which is relatively
easy to automatize, enables us to 
go one order further in the perturbation theory, with much less
computational effort than Feynman diagrammatic expansions would require.

In the weak noise limit the kernel is sharply peaked, so it
makes sense to expand it
in terms of the Dirac delta function and
its derivatives:
\beq
	\delta_\sigma(y)
	= 
	\sum_{m=0}^{\infty} {a_m \sigma^m \over m!} \, \delta^{(m)}(y) 
	=
	\delta(y) + 
	a_2 {\sigma^2 \over 2} \delta^{(2)}(y) +
	a_3 {\sigma^3 \over 6} \delta^{(3)}(y) + \dots
	\,.
\label{delSigExp}
\eeq
where
\[
	\delta^{(k)}(y) = {\pde^k \over \pde y^k} \delta(y)
	\,,
\]
and the coefficients $a_m$ depend on the choice of the kernel.
We have omitted the $\delta^{(1)}(y)$ term in the above because 
in our applications we shall impose
the saddle-point condition, that is, 
we shift $f$ by a constant to ensure that the noise peak corresponds
to $y=0$, so $\delta_\sigma^{'}(0)=0$.
For example, if $\delta_\sigma(y)$ is a Gaussian kernel,
it can be expanded as
\bea
	\delta_\sigma(y)
	&=& 
	{1 \over \sqrt{2 \pi \sigma^2}} e^{-{y^2/2\sigma^2} }
	=
	\sum_{n=0}^{\infty}
	\frac{\sigma^{2n}}{n!2^n} \delta^{(2n)}(y)
	\continue
	&=& 
	\delta(y) + {\sigma^2 \over 2} \delta^{(2)}(y) 
	 + {\sigma^4 \over 8} \delta^{(4)}(y) + \cdots
	\,.
\label{delGaussExp}
\eea

\section{Stochastic trace formula}

We start our computation of the weak noise corrections to the 
spectrum of $\Lnoise{}$ by calculating the trace of the $n$th iterate of
the stochastic evolution operator $\Lnoise{}$
for a one-dimensional analytic
map $f(x)$ with additive noise $\sigma$.  
This trace is an $n$-dimensional
integral on $n$ points along a discrete periodic chain, 
so $x$ becomes an $n$-vector $x_a$ with indices $a,b,\ldots$
ranging from $0$ to $n$$-$$1$ 
in a cyclic fashion
\bea
\tr{\Lnoise{n}} &=& \int 
	\prod_{a=0}^{n-1} dx_a \, \delta_\sigma(y_a)
	\continue
y_a(x) &=& f(x_{a})  - x_{a+1}\,,\qquad x_n  =  x_0 
\,.
\label{FatIntDef}
\eea

To the order $\sigma^3$ the composition is simple: 
all compositions but one 
can be made, resulting in 
\bea
	\tr {\Lnoise{\cl{}}}
	& = & \int dx \, \delta(f^\cl{}(x)-x) + 
		{{a_2}\over 2}\sigma^2
		\int dx_1 \dots dx_\cl{} 
        	\ceq
		\sum_{a=0}^{\cl{}-1}
		\delta(f(x_{\cl{}})-x_1) \dots
		\delta^{(2)}(f(x_{a})-x_{a+1}) \dots
		\delta(f(x_{1})-x_2)
		\,+\, \cdots
	\continue
	&=& \tr {\Lop^{\cl{}}}
		+ {a_2\over 2}{\sigma^2}
		\sum_{a=0}^{\cl{}-1}
		\int dx \, 
		\delta^{(2)}(f^\cl{}(x_a)-x_a)
	\ceq
		+ {a_3\over 3!}{\sigma^3}
		\sum_{a=0}^{\cl{}-1}
		\int dx \, 
		\delta^{(3)}(f^\cl{}(x_a)-x_a)
	 \,+\, O(\sigma^4)
	\,.
\label{compose5}
\eea
At fourth order we get contributions from
$\delta^{(4)}$, as well as the two-point contributions
\bea
	\tr {\Lnoise{\cl{}}}
	& = & (\cdots) 
	+ {a_4\over 4!}{\sigma^4}
                \sum_{a=0}^{\cl{}-1}
		\int dx \,
                \delta^{(4)}(f^\cl{}(x_a)-x_a)
        \ceq
	+ {a^2_2\over 4}\sigma^4
	\sum_{a<b}\int dx_a dx_b \,
	\delta^{(2)}(f^j(x_a)-x_b)
	\delta^{(2)}(f^k(x_b)-x_a)
	\,,
\label{4htOrd}
\eea
where $j$ is the number of steps from points $a$ to $b$ on the cycle,
and $k$ is the number of step from $b$ to $a$, so that $j+k=n$.

If the map is smooth, the periodic points of given finite
period $n$ are isolated and the noise broadening $\sigma$
sufficiently small so that they remain separated, the dominant
contributions come from neighborhoods of periodic points;
in the
{\em saddlepoint approximation} the trace \refeq{FatIntDef} is given by
\beq
\tr{\Lnoise{n}} \longrightarrow \sum_{x_c\inFix{n}} e^{W_c}
\,,
\ee{SptSum}
As traces are cyclic, 
$e^{W_c}$ is the same
for all periodic points in a given cycle, independent of the choice
of the starting point $x_c$.
Hence it is customary to rewrite this sum in terms of 
prime cycles and their repeats,
\beq
\left. \tr{\Lnoise{n}} \right|_{\mbox{\footnotesize saddles}}
  = \sum_p \cl{p} \sum_{r=1}^\infty  e^{W_{p^r}}
\,,
\ee{SptSum1}
where $p^r$ labels the $r$th repeat of prime cycle $p$.

\section{Trace evaluated at a fixed point to all orders}
\label{s:TraceFixPoint}

We now derive the perturbative expansion 
for a fixed point ($n=1$) to {\em all orders} in $\sigma$.
A fixed point and its repeats are of particular interest having
the same interaction at every site, as does the usual field theory.
What we do here is to formulate (and partially solve, in the
sense of determing a few orders of the exact perturbation theory
expansion) the field theory on finite periodic 1-dimensional 
discrete chains.

Defining
$
y = f(x) -x
\,,
$
we can write the fixed point trace as
\beq
	\tr{\Lnoise{}}
	=
	\averx \delta_\sigma(f(x)-x)
	= \int dy \, {1\over \left|y'(x)\right|}  \delta_\sigma(y)
	\,.
\label{(10)}
\eeq
Expanding the kernel $\delta_\sigma(y)$ 
as in \refeq{delSigExp} and integrating by parts, we see that
all is well if we know the $d/dx$ derivatives of $1/y'(x)$. Replacing
\[
{d \over d y} \to {\pde x \over \pde y}{d \over d x}
	=  {1\over y'(x)} {d \over d x}
\]
we obtain in the saddlepoint approximation contributions to each 
fixed point of $f$ evaluated recursively as derivatives
of $1/y'(x)$
\[
\int dx \, \delta^{(k)}(y)
	=
	\sum_{x : y(x)=0}
(-1)^k {d^k \over d y^k} {1 \over \left|y'(x)\right|}
	=
	\sum_{x : y(x)=0}
\left(- \frac{1}{y'(x)}{d \over d x} \right)^k\frac{1}{|y'(x)|}
\,.
\nnu
\]
The $d/dy$ derivatives of $1/y'$ are related to the $d/dx$ 
derivatives of the map $f$ by
\bea
{1 \over y'} &=&  {1 \over f'(x)-1} 
		\continue
{d \over d y} {1 \over y'} &=&  
	-{ f'' \over (y')^3}  
		\continue
{d^2 \over d y^2} {1 \over y'} &=&  
	- { f''' \over (y')^4}  
	+ 3 { (f'')^2 \over (y')^5}  
		\label{F_3rde}
		\\
{d^3 \over d y^3} {1 \over y'} &=&  
	-{ f{''''} \over (y')^5}  
	+ { 10 f'' f''' \over (y')^6}  
	- { 15 (f'')^3 \over (y')^7}  
\nnu
\eea
Here, and also later, we have relegated 
the fourth and fifth
derivatives \refeq{F_5rde} to the Appendix. 
For example, for the second derivative of the
delta function we have
\beq
	\int dx \, \delta^{(2)}\left(y\right)
	=
	\sum_{x : y(x)=0}
		{1 \over |y'|}
	\left(
		3 {(y'')^2 \over (y')^4}
        	-
		{y''' \over (y')^3}
	\right)
	\,,
\ee{delta_2}
where the sum is over all fixed points $f(x)=x$.
In general, for $n\geq 1$
\beq
	(-1)^n {d^n \over d y^n} {1 \over |y'|} =
	{1 \over |y'|}
\sum_{\{k_\ell\}}
\frac{(2n-k_1)!}{(-y')^{2n-k_1}}
\prod_{\ell \geq2}
\frac{{f^{(\ell)}}^{k_\ell}}{(\ell !)^{k_\ell}k_\ell !}
\,.
\eeq
where $f^{(\ell)}$ is the $\ell$-{th} derivative and the sum is over all
sets $\{k_\ell\}$ satisfying $\ell\geq 1$, $k_\ell\geq 0$, $\sum k_\ell=n$, and
$\sum \ell k_\ell=2n$, that is, all partitions of $2n$ into $n$ terms.  The
product contains only the finitely many nonzero $k_\ell$ for $\ell\geq 2$.
This formula may be proved in a staightforward manner by induction.
It may be simplified substantially by defining a generating function
\bea
{\cal F}(z,x)
	&=&
	\sum_{n=1}^{\infty}z^{n}
	\left. 
	{d^{n-1} \over d y^{n-1}} {1 \over y'(x)}
	\right|_{x:y(x)=0}
	\continue
	&=&
	\exp\left(-z^{-1}\sum_{j=2}^{\infty}
	\frac{f^{(n)}z^n\partial_{y'}^n} {n!}\right) \frac{z}{y'}
\,,
\eea
where $\partial_{y'}=\partial/\partial y'$ acts only on $y'$.  The
sum in the exponential is formally a Taylor series expansion of
\beq
{\cal F}(z,x)=\exp\left\{-z^{-1}\left[f(x-\partial_u)+\Lambda\partial_u
-x\right]\right\}
\frac{1}{u}
\,,
\eeq
with $u=y^{'}/z$, $f'(x)=\Lambda$ and $f(x)=x$.

\section{Smooth conjugacies}
\label{scfpo}

The next step 
injects into field theory a method
standard in the analysis of fixed points and 
construction of normal forms for bifurcations,
see refs.~\cite{Katok95}--\cite{MiraGum80}.
The idea is to perform a smooth
nonlinear change of coordinates that flattens out the vicinity
of a fixed point and makes the map {\em linear} in an open neighborhood.
This can be implemented
only for an isolated nondegenerate fixed point (otherwise
higher terms will contribute to the normal form expansion around the
point), and only in a finite neigborhood of a point, as the 
conjugating function in general has a finite radius of convergence.
Later we extend the method to periodic orbits, which are fixed points
of the $n$th iterated map.

\subsection{Fixed points}

Let the fixed point of $f(x)$ be $x=0$
and the stability of that point be $\Lambda = f'(0)$. 
If $|\Lambda|\neq 1$,
there exists a smooth conjugation $h(x)$ satisfying $h(0)=0$ such that: 
\beq
	f(x) = h(\Lambda h^{-1}(x))
\,.
\label{EqConj}
\eeq
In several dimensions, $\Lambda$ is replaced by the Jacobian matrix,
and one has to check that its eigenvalues are non-resonant, that is,
there is no integer linear relation between their logarithms. 
If $h(x)$ is a conjugation, so is any scaling $h(\alpha x)$ of the function
for a real number $\alpha$. Hence
the value of $h'(0)$ is not determined by the functional equation; we  
shall set $h'(0)=1$.

To compute the conjugation $h$ we use the functional equation 
$ h(\Lambda x)=f(h(x))$ and the expansions
\bea
	f(x) 
	&=&
	\Lambda x + x^2 f_2 + x^3f_3 + \dots
	\continue
	h(x) 
	&=&
	x + x^2 h_2 + x^3 h_3 + \dots
	\;\,.
\label{f-hExps}
\eea
In the present context absorbing the factorials into the definition of
expansion coefficients
turns out to be more convenient than the standard Taylor expansion.
Equating recursively coefficients in expansions
\bea
	h(\Lambda u) - \Lambda h(u) &=& 
		\sum_{n=2}^\infty f_m \left(h(u)\right)^m
		\continue
\sum_{n=2}^\infty (\Lambda^n-\Lambda) h_n u^n
	&=&
\sum_{m=2}^\infty f_m  u^m \left(1+ \sum_{k=2}^\infty h_k u^{k-1} \right)^m
	\label{EqConjExp}
\eea
yields
\beq
h_2 = \frac{f_2}{ \Lambda (\Lambda - 1) }
	\,,\qquad
h_3 =
\frac{2 f_2^2 + \Lambda  ( \Lambda -1  )   f_3}
   { \Lambda^2 
        ( \Lambda -1 ) 
        ( \Lambda^2 -1 ) 
   }
\,, \qquad\cdots
\ee{h(2)}
Noting that the left-hand-side of \refeq{EqConjExp}
generates  denominators in
\refeq{h(2)} 
are the same as those appearing in the Euler formula
\begin{eqnarray*}
\prod_{k=0}^{\infty} (1+tu^k)
&=&
1 + \frac{t}{1-u} + \frac{t^2u}{(1-u)(1-u^2)}
  + \frac{t^3u^3}{(1-u)(1-u^2)(1-u^3)}   \cdots\\
&=&
\sum_{k=0}^{\infty} t^k \frac{u^\frac{k(k-1)}{2}}{(1-u)\cdots (1-u^k)}
\,, \quad \quad |u|<1
\,,
\end{eqnarray*}
we find it convenient to factorize $h_n$ as
\[
h_n = {b_n \over D_n}
\,,\qquad
D_n = \left(1-{1 \over \Lambda}\right)\left(1-{1 \over \Lambda^2}\right)\cdots \left(1-{1 \over \Lambda^{n-1}}\right)
	\Lambda^{\frac{(n+2)(n-1)}{2}}
\,.
\]
Computer algebra then yields
{\small
\bea
b_2 &=& f_2
\continue
b_3 &=&
    2f_2^2 + \Lambda \left( \Lambda -1  \right)  f_3
\label{h(5)}\\
b_4 &=&
	(5 + \Lambda) f_2^3 
    - \Lambda  (5 - 2\Lambda  - 3\Lambda^2 ) f_2f_3 
	+ \Lambda^2 (\Lambda - 1) (\Lambda^2 - 1) f_4
\nonumber
\,.
\eea
} 
The expressions 
\refeq{h(5)a}
for $b_5$ and $b_6$ are given in the Appendix.

\subsection{Longer cycles}

Now that we have constructed the conjugation function for a
fixed point, we turn to the problem of constructing it for
periodic orbits. Each point around the cycle has a
differently distorted neighborhood,
with differing second and higher derivatives,
so we need to compute a conjugation function $h_a$ at each 
cycle point $x_a$.
We expand the map $f$ around each cycle point along the cycle,
\[
y_a(\phi)=f_a(\phi)-x_{a+1}=\phi f_{a,1}+ \phi^2 f_{a,2}+\ldots
\]
where $x_a$ is a point on the cycle, $f_a(\phi)=f(x_a+\phi)$
is centered on the deterministic orbit,
and the index $k$ in $f_{a,k}$ refers to
the $k$th order in the (modified) Taylor expansion
\refeq{f-hExps}.

For a periodic orbit the conjugation formula
\refeq{EqConj}
generalizes to 
\[
f_a(\phi)=h_{a+1}(f'_a(0)h_a^{-1}(\phi))
\]
at each point.
The conjugationg functions $h_a$ are obtained in the same way as before, by
equating coefficients of Taylor series, and assuming that
the cycle stability is not marginal, $|\Lambda|\neq 1$.
The explicit expressions for $h_a$ in terms of $f$ are 
obtained by iterating around the whole cycle,
\beq
f^n(x_a+\phi)=h_a(\Lambda h_a^{-1}(\phi))+x_a
\,,
\ee{hCycle}
so each $h_a$ function is given by the derivatives given
in the previous section acting on $f^n$, evaluated at each 
cycle point $a$ (recursive formulas for evaluation of such derivatives are
given in Appendix A of \refref{noisy_Fred}).
Again we have the freedom to set $h_a'(0)=1$ for all $a$.

We shall also find it convenient to
define partial stabilities along cycle laps
\begin{eqnarray*}
f^j(x_a+\phi)&=&x_b+h_b(\Lambda_j h_a^{-1}(\phi))\\
f^k(x_b+\phi)&=&x_a+h_a(\Lambda_k h_b^{-1}(\phi))
\end{eqnarray*}
where
\[
\Lambda_j=\prod_{c=a}^{b-1}f'(x_c),\quad
\Lambda_k=\prod_{c=b}^{a-1}f'(x_c)
\]
and $\Lambda_j\Lambda_k=\Lambda$.

At this point we note that while in the deterministic case the cycle
weight is given by the cycle stability, a quantity invariant under all
smooth coordinate transfomations, the higher order corrections are not
invariant. The noise is defined with respect to a particular coordinate
system, and it has no invariant meaning.

\section{Trace formula for repeats of periodic orbits}

What is gained by rewriting the perturbation expansion for
$f$ in terms of the equally messy perturbation expansion for the
conjugacy function $h$?
Once the
neighborhood of a fixed point is linearized, the repeats of it
are trivialized; 
from the conjugation formula \refeq{EqConj}
one can compute the derivatives of a function composed with itself $r$ times:
\[
	f^r(x)=h(\Lambda^r h^{-1}(x))
	\,.
\]
One can already discern the form of the expansion for arbitrary
repeats; the answer will depend on the conjugacy function $h(x)$
computed for a {\em single} repeat, and all the dependence on the
repeat number will be carried by factors polynomial in $\Lambda^r$,
a result that emerged as a surprise in the Feynman diagrammatic
approach\rf{noisy_Fred}.

\subsection{Repeats of fixed points}

The above observation enables us to move from a field theory constructed at a
single point to a field theory which is translationally invariant on
a periodic chain of arbitrary length $r$.
The first and the second derivatives are
\bea
	{d \over d x} {f^r}(x)
	&=&
	h'(\Lambda^r h^{-1}(x))\Lambda^r\frac{1}{h'(h^{-1}(x))}
	\continue
	{d^2 \over d x^2} f^{r}(x)
	&=&
	\frac{\Lambda^{2r} h''(\Lambda^r h^{-1}(x))}{h'(h^{-1}(x))^2}
	- 
	\frac
		{\Lambda^{r} h''(h^{-1}(x)) h'(\Lambda^r h^{-1}(x))}
		{h'(h^{-1}(x))^3}
	\,.
\label{f(1)f(2)}
\eea
The third derivative is too long to write down here.
Evaluation at the fixed point $x=0$ yields
\bea
	{d \over d x} {f^r}(0)
	&=&
	\Lambda^r
	\,,\qquad
	{d^2 \over d x^2} f^{r}(0)
	=
	\Lambda^r (\Lambda^{r} - 1) {h^{(2)}}
	\continue
	{d^3 \over d x^3} f^{r}(0)
	&=&
	\Lambda^r (\Lambda^{2r} - 1)  h^{(3)}
	  -
	3 \Lambda^r (\Lambda^r - 1) {h^{(2)} }^2
	\,\, \cdots \,.
\label{f(1)f(2)fix}
\eea
(For brevity we shall often replace $h^{(r)}(0) \to h^{(r)}=r!h_r$,
$f^{(r)}(0) \to f^{(r)}=r!f_r$ throughout, where the superscript
$^{(r)}$ labels the $r$th coefficient in the standard Taylor expansion.)
In general, the $k$th derivative of a function composed with itself $r$ times 
is a polynomial in $\Lambda^r$ of degree $k$
\[
	{d^k \over d x^k} f^{r}(0)
	=
	\sum_{j=1}^{k} c_j \Lambda^{j r}
	\,.
\]
The coefficients $c_j$ are in turn expressed in terms of the first $k$
derivatives $h^{(j)}$ of the conjugation function. 

Now we can reevaluate the fixed-point derivatives of 
\refsect{s:TraceFixPoint} for the case of $r$th repeat of a
fixed point
\beq
y(x) =  f^{\cl{p}r}(x) - x
\ee{g(x)}
in terms of the conjugating function $h$
by substituting \refeq{f(1)f(2)fix} into \refeq{F_3rde}:
{\small
\bea
\left.\frac{1}{3!}\frac{\partial^2}{\partial y^2}
\frac{1}{y'(x)}\right|_{x=0}
	&=&
	{\frac{{{\Lambda }^r}\left( 1 + {{\Lambda }^r} \right) 
	}{{{\left( \Lambda^r -1 \right)
          }^3}}}
      \left(2h_2^2 -  h_3 \right)
\label{der(5)} \\
	\left.\frac{1}{4!}\frac{\partial^3}{\partial y^3}
	\frac{1}{y'(x)}\right|_{x=0}
	&=&
	-5\Lambda^r{ {( \Lambda^r + 1 )^2}
	    \over
            (\Lambda^r - 1)^4 }h_2^3 
        + \Lambda^r { 5\Lambda^{2r} + 8\Lambda^r + 5
	    	      \over
            	      (\Lambda^r - 1)^4 } h_2h_3 
		\ceq
	~~~~~~~~~~~~~~~~~~~~~~
         - \Lambda^r { \Lambda^{2r} + \Lambda^r + 1
                      \over
                      (\Lambda^r - 1)^4 } h_4
\label{der(3)}
\eea
} 
The 4th and 5th order derivatives 
\refeq{der(5)a}
are relegated to the Appendix.

\subsection{Repeats of periodic orbits}

The above fixed-point formulas carry over to the case of 
general periodic orbits, if $f$ is identified with the $n$th
iterated map at a particular point along the cycle $a$ from which we
define $h_a$ as in \refeq{hCycle}.  The second
derivative required for the $\sigma^2$ correction
follows immediately, with \refeq{der(5)} evaluated for each cycle point 
in the prime cycle trace \refeq{compose5}.
It is easily checked that this second derivative 
\refeq{der(5)} has exactly
the same form as what we obtained in a much more involved manner
in \refref{noisy_Fred}.
The somewhat more complicated sum of Feynman diagrams is here replaced 
by \refeq{der(5)}, the conjugation function determined from the
iterated map, with $h$ carrying all dependence on the 
higher derivatives 
of the original map.

The factorisation is not quite as simple for
higher orders in the trace formula. 
For a general non-Gaussian case at order $\sigma^3$
we have three terms from 
\refeq{der(3)}, each of which may be resummed
separately.  The calculation is entirely analogous to the second order
calculation in \refref{noisy_Fred}. Powers of $\Lambda^r$ are moved
around in order to express
\refeq{der(3)} in terms of $\left|\Lambda\right|^{-r}<1$ and generate
convergent geometric series; each of the three terms is expanded using
identities such as
\beq
3\frac{(1+x)^2}{(1-x)^4}=\sum_{k=0}^{\infty}k(1+2k^2)x^k
\,,
\ee{sumA}
so that repeats $r$ can be resummed. 
(The remaining identities \refeq{sum0}--\refeq{sumC}
used here are relegated to the Appendix.)
This yields the saddlepoint approximation to
the spectral determinant in product form up to order $\sigma^3$
\beq
\det(1-z{\cal L}_\sigma) = \prod_p\prod_{k=0}^{\infty}(1-t_{p,k})
\ee{zetasigma3}
\bea
t_{p,k}&=&\frac{z^{n_p}}{|\Lambda_p|}
		\frac{1}{\Lambda_p^k}
\exp{\left(\frac{1}{2}a_2\sigma^2w^{(2)}_{p,k}
         +\frac{1}{3!}a_3\sigma^3w^{(3)}_{p,k}+O(\sigma^4)\right)}
	\continue
w^{(2)}_{p,k}&=&(k+1)^2\sum_a\left(2h_a^{(2)^2}-h_a^{(3)}\right)
	\continue
w^{(3)}_{p,k}&=&-\frac{k(1+2k^2)}{3}\sum_a h_a^{(2)^3}
	+(1+k)(5+6k+3k^2)\sum_a h_a^{(2)}h_a^{(3)}
	\ceq
	-\frac{(1+k)(2+2k+k^2)}{2}\sum_a h_a^{(4)}
\nnu
\eea
The meaning of this factorization is that in the saddlepoint
approximation the spectral determinant is composed
of local spectra evaluated for
each prime cycle separately, with the index $k$ labelling
the $k$th local eigenvalue.

From $\sigma^4$ onward further terms come into play,
as we now describe.

\section{Two-point integrals}
\label{s-Two-p-int}

In general, a saddlepoint at each cycle point 
may be expanded in second and higher derivatives
(\ref{delSigExp}), with the trace to $n$th order receiving
a single $\delta^{(n)}$ integral contribution as
in \refeq{compose5}.
At fourth order we need to also include the two-point integral
contribution \refeq{4htOrd}.
Setting $y_j=f^j(x_a)-x_b$, $y_k=f^k(x_b)-x_a$, and integrating by parts,
we obtain for each pair of points along the cycle
\[
\int \frac{dy_j dy_k}{|f^{j'}(x_a)f^{k'}(x_b)-1|}\delta^{(2)}(y_j)
\delta^{(2)}(y_k)
	=
\frac{\partial^2}{\partial y_j^2}
\frac{\partial^2}{\partial y_k^2}
\frac{1}{|f^{j'}(x_a) f^{k'}(x_b)-1|}
\,.
\]
This expression looks innocent enough, but its evaluation requires
attention to some subtle points, and more algebra than 
a human would be willing to handle.  
Performing the required derivatives taxes the ability of general purpose
symbolic algebra packages, so the results in this section are obtained
with a program written in C. 

Consider evaluating partial derivatives of a function $q$ with respect to,
say $y_j$ keeping $y_k$ constant in terms of $x_a$ and $x_b$ derivatives.
The solution is to consider an arbitrary infinitesimal transformation
which respects the $dy_k=0$ constraint:
\begin{eqnarray*}
dq&=&\frac{\partial q}{\partial x_a}dx_a
+\frac{\partial q}{\partial x_b}dx_b\\
dy_j&=&\frac{\partial y_j}{\partial x_a}dx_a
+\frac{\partial y_j}{\partial x_b}dx_b\\
dy_k&=&\frac{\partial y_k}{\partial x_a}dx_a
+\frac{\partial y_k}{\partial x_b}dx_b=0
\end{eqnarray*}
These equations can then be solved for the ratio of $dq$ and $dy_j$,
\[
\frac{\partial q}{\partial y_j}
=
	{
\frac{\partial q}{\partial x_a}\frac{\partial y_k}{\partial x_b}
-\frac{\partial q}{\partial x_b}\frac{\partial y_k}{\partial x_a}
	 \over 
\frac{\partial y_j}{\partial x_a}\frac{\partial y_k}{\partial x_b}
-\frac{\partial y_j}{\partial x_b}\frac{\partial y_k}{\partial x_a}
	}
\]
and similarly for $\partial q/\partial y_k$.

Sixth order corrections require a three point integral, which we leave
as an exercise for the reader.

If for some reason the noise kernel contains a linear $a_1 \sigma$ term
in the expansion \refeq{delSigExp}, for
example if the evolution operator is weighted, and the shift that this
causes to the saddlepoint is not taken into account, the relevant second
order correction includes the term
\begin{eqnarray*}
& &
\frac{\partial}{\partial y_j}
\frac{\partial}{\partial y_k}
\frac{1}{|(\partial f^{j}(x_a)/\partial x_a)
( \partial f^{k}(x_b)/\partial x_b)-1|}
\\
&=&\frac{\Lambda}{(\Lambda-1)^3}
\left[(\Lambda+1)h_a^{(2)}h_b^{(2)}+2\Lambda_k h_a^{(2)^2}
+2\Lambda_j h_b^{(2)^2}-\Lambda_k h_a^{(3)}-\Lambda_j h_b^{(3)}\right]
\end{eqnarray*}
This we have written down just to illustrate the form
of a 2-point term. In our application the saddlepoint condition
sets $a_1=0$,
and what is really required is the  
two-point fourth order correction \refeq{2p4}
and (in case of non-Gaussian noise) the two-point fifth order correction
\refeq{2p5th}.
The corresponding expressions are too cumbersome for the main text, so we
relegate them to Appendix.

There are some convenient cross-checks
on the algebra. The results reduce to those
of \refeq{der(3)} and  \refeq{der(5)a} for single
repeat $r=1$, by setting $h_a=h_b=h$,
$\Lambda_j=\Lambda$, $\Lambda_k=1$, and remebering
that in order
to retain integer coefficients, we have here replaced power series
coefficients $h_r$ by the Taylor expansion
derivatives $h^{(r)}=r!h_r$.  This reduction to the one-point case
also means that for the Gaussian case ($a_2=1$, $a_4=3$), 
expression \refeq{4htOrd} can be written as an unrestricted
double sum,
\[
\frac{\sigma^4}{8}\sum_{x\in{\rm Fix}f^n}\sum_{ab}
\frac{\partial^2}{\partial y_j^2}
\frac{\partial^2}{\partial y_k^2}
\frac{1}{|(\partial f^{j}(x_a)/\partial x_a)
( \partial f^{k}(x_b)/\partial x_b)-1|}
\]
where it is understood that $j=n$ and $k=0$ when $a=b$.

Now the stage is set to return to the investigations of
\refref{noisy_Fred}, where the $\sigma^2$ term was evaluated
anlytically, while the $\sigma^4$ term was only estimated
from the numerically computed leading eigenvalue. 

\section{Numerical tests}

Here we continue the calculations of Sect.~5 of \refref{noisy_Fred},
where more details and discussion may be found.  
We test our perturbative expansion on the repeller
of the 1-dimensional
map $f(x)=20[(1/2)^4-((1/2)-x)^4]$.
This repeller is a nice example of an ``Axiom A'' expanding
system of bounded nonlinearity and complete binary symbolic dynamics, 
for which the deterministic evolution operator eigenvalues
converge super-exponentially with the cycle length.
We compute the leading eigenvalue of the evolution operator 
(the repeller escape rate) in the presence of Gaussian noise, using two
complementary approaches.  The perturbative result in terms
of periodic orbits and the weak noise corrections
is compared to a numerical eigenvalue obtained in \refref{noisy_Fred} 
by approximating the dynamics by a finite matrix. In the preceding
paper we compared the numerical eigenvalue with the $\sigma^2$ result
and estimated the coefficient of $\sigma^4$ to be approximately 38.  Here we
compute the order $\sigma^4$ coefficient 
to 14 digits accuracy, and estimate the $\sigma^6$ term.
In the Feynman diagram language, the  $\sigma^4$ contribution 
is a ``2-loop'' calculation, albeit one of relatively simple kind
where space integrals are replaced by discrete sums over a 
finite periodic chains.
The conjugation functions \refeq{hCycle} at each point
around the cycle are obtained by a recursive evaluation of~\refeq{EqConjExp},
and then substituted 
into~\refeq{der(5)} 
for the second order and~\refeq{2p5th}
for the fourth order.  The expansion for the spectral determinant is obtained
by differentiating $\ln\det(1-z{\cal L}_\sigma)=\tr\ln(1-z{\cal L}_\sigma)$
with respect to $z$, multiplying through by the determinant, and equating
coefficients order by order in $z$ and $\sigma$, as in \refref{noisy_Fred}.
That is, we define coefficients
\[
\tr{\cal L}_\sigma^n=\sum_{j=0}^{\infty}C_{n,j}\sigma^j
	\,,\quad\quad
\det(1-z{\cal L}_\sigma)=1-\sum_{n=1}^{\infty}\sum_{j=0}^{\infty}
Q_{n,j}z^n\sigma^j\label{Fexp}
\,,
\]
and obtain the cumulants $Q_{n,j}$ recursively as
\bea
Q_{n,0}&=&\frac{1}{n}\left[C_{n,0}-\sum_{k=1}^{n-1}Q_{k,0}C_{n-k,0}\right]
	\continue
Q_{n,2}&=&\frac{1}{n}\left[C_{n,2}-\sum_{k=1}^{n-1}\left(Q_{k,2}C_{n-k,0}+
Q_{k,0}C_{n-k,2}\right)\right]
	\continue
Q_{n,4}&=&\frac{1}{n}\left[C_{n,4}-\sum_{k=1}^{n-1}\left(Q_{k,4}C_{n-k,0}+
Q_{k,2}C_{n-k,2}+Q_{k,0}C_{n-k,4}\right)\right]
\,,
\nnu
\eea
where it is understood that the sums do not contribute when $n=1$.

The noiseless, zeroth order eigenvalue equation
$\det(1-z{\Lop}_0)=0$ is solved by Newton's
method to find the leading eigenvalue $\nu_0=z^{-1}$ at $\sigma=0$, and
the higher order equations give the noise corrections 
$\nu(\sigma)=\nu_0+\nu_2\sigma^2+\nu_4\sigma^4+O(\sigma^6)$ in terms
of $\nu_0$ and the expansion of the determinant. 
Expanding the spectral determinant order by order in $z$ and $\sigma$
we find
\begin{equation}
\det(1-z{\cal L}_\sigma)=F-F_{10}z-F_{02}\sigma^2-F_{20}z^2-F_{12}z\sigma^2
-F_{04}\sigma^4-\ldots
\end{equation}
where the expansion coefficients are
\bea
F&=&1-\sum_{m=1}^n\frac{Q_{m,0}}{\nu_0^m}
\,,\quad
F_{10} = \sum_{m=1}^n\frac{mQ_{m,0}}{\nu_0^{m-1}}
\,,\quad
F_{02} = \sum_{m=1}^n\frac{Q_{m,2}}{\nu_0^m}
	\continue
F_{20}&=&\sum_{m=2}^n\frac{m(m-1)Q_{m,0}}{2\nu_0^{m-2}}
\,,\quad
F_{12} = \sum_{m=1}^n\frac{mQ_{m,2}}{2\nu_0^{m-1}}
\,,\quad
F_{04} = \sum_{m=1}^n\frac{Q_{m,4}}{\nu_0^m}
\nnu
\eea
are obtained from derivatives of (\ref{Fexp}).  Finally we expand
$\nu=z^{-1}$ in powers of $\sigma^2$ and equate coefficients of powers
of $\sigma$ to obtain
\[
\nu_2 = \frac{F_{02}\nu_0^2}{F_{10}}
\,,\quad
\nu_4 = \frac{F_{20}F_{02}^2-2F_{12}F_{10}F_{02}+F_{04}F_{10}^2+F_{10}F_{02}^2\nu_0}
{F_{10}^3}\nu_0^2
\]

\begin{figure}
\centering{\includegraphics[width=0.90\textwidth]{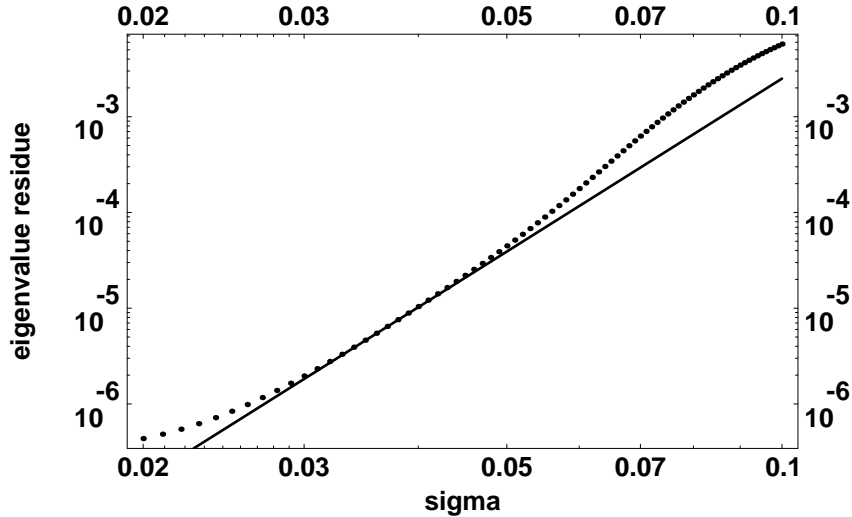}}
\caption{
Numerically computed eigenvalue minus known terms, that is,
$\nu(\sigma)-\nu_0-\sigma^2\nu_2-\sigma^4\nu_4$ (dots), together
with estimated next term $2700\sigma^6$ (solid line).  Deviations
occur at small $\sigma$ due to discretization errors in the numerical
eigenvalue of a few times $10^{-7}$, and at large $\sigma$ due to further
omitted terms ($\sigma^8\ldots$).
}
\label{FigResidue} 
\end{figure}

The perturbative corrections to the leading eigenvalue
(escape rate) of the weak-noise evolution operator
are given in table \ref{tab},
showing super-exponential convergence with the truncation 
cycle length $n$.  The super-exponential convergence
has been proven for the deterministic,
$\nu_0$ part of the eigenvalue\rf{grothi,Rugh92}, but has not been
studied for noisy kernels.
It is seen that a good first approximation
is obtained already at $n=2$, using only 3 prime cycles,
and $n=6$ (23 prime cycles in all) is in this example
sufficient to exhaust the limits of double precision arithmetic.
The exact value of $\nu_4 = 36.358\ldots$ is encouragingly close
to  our previous numerical estimate\rf{noisy_Fred}  of 38.
 As in the preceding paper\rf{noisy_Fred}, we subtract the known terms
in the expansion from the numerically evaluated eigenvalue, and obtain
a good fit to the next term, approximately $2700\sigma^6$, 
see \reffig{FigResidue}.

\begin{table}
\begin{tabular}{clll}\hline
$n$&$\nu_0$&$\nu_2$&$\nu_4$\\\hline
1&0.308&0.42&~2.2\\
2&0.37140&1.422&32.97\\
3&0.3711096&1.43555&36.326\\
4&0.371110995255&1.435811262&36.3583777\\
5&0.371110995234863&1.43581124819737&36.35837123374\\
6&0.371110995234863&1.43581124819749&36.358371233836\\\hline
\end{tabular}
\caption{Significant digits of the leading deterministic eigenvalue
and its $\sigma^2$ and $\sigma^4$ coefficients, calculated from the
spectral determinant as a function of the cycle truncation length $n$.
Note the super-exponential convergence of all coefficients 
(the $n=6$ result is here limited by the machine precision).
\label{tab}}
\end{table}

\section{Summary and outlook}

We have formulated a perturbation theory of
stochastic trace formulas based on smooth
nonlinear field transformations
around infinitely many chaotic saddle points 
(unstable periodic orbits). In contrast to
previous perturbative expansions around
vacua and instanton solutions, the location
and local properties of each saddlepoint must
be found numerically.  In addition, every
interaction term depends on the position,
hence also the classical periodic solution at
which it is evaluated.

Even though in the model
calculation we have chosen a Gaussian one-step noise,
the accumulated noise along a trajectory
is distorted by the nonlinear flow, and orbit-by-orbit
the noise corrections are not Gaussian.  Gaussian noise
has thus no privilaged role in nonlinear dynamical systems.

The key idea is this: Instead of separating the action into a
quadratic and ``interaction'' parts, we first perform a nonlinear 
field transformation (``smooth conjugation'') which turns the
saddle point into an exact
quadratic form.  The price one pays for this is the Jacobian of
the nonlinear field transformation -
but it turns out
that the perturbation expansion of this Jacobian 
{\em in terms of the conjugating function} is order-by-order
considerably more
compact than the Feyman-diagrammatic expansion.

We have resummed repeats of prime cycles to third order in the noise
strength, carried out the numerical tests to fourth order,
given the trace formula for general periodic orbits to fifth order,
and for a fixed point to all orders.

The rapid rate of increase of the numerical coefficients confirms the
expectation that the series is
asymptotic, and is to be used with caution, as long as
no sophisticated summation beyond all orders is
implemented.

The smooth conjugacy
method of perturbation expansions can be extended to the case of
higher dimensions and continuous time dynamics (stochastic flows),
but the main
interest comes from the observation that we have a new method
of evaluating perturbative corrections to saddlepoints of
path integrals.
In quantum mechanics and field theory
the perturbative corrections do matter, and the method might
have applications there, in particular
to the $\hbar$ expansion of semiclassical
periodic orbit theory.
If efficient methods are found for
computing numerical periodic solutions of spatially extended systems,
the method might apply to the field theory as well.

\appendix
\section{Appendix: Some algebra}

We collect here some of the formulas used in our calculations,
but too long for the patience of a casual reader.

Evaluation of the 
trace of a fixed point (\ref{F_3rde}) continued to 4th and 5th order:
\bea
{d^4 \over d y^4} {1 \over y'} &=&  
        - { f^{(5)} \over (y')^6}
        + { 15 f'' f'''' \over (y')^7}
        + { 10 (f''')^2 \over (y')^7}
        - { 105 (f'')^2 f''' \over (y')^8}
        + { 105 (f'')^4 \over (y')^9}
                \continue
{d^5 \over d y^5} {1 \over y'} &=&
        -\frac{f^{(6)}}{(y')^7}
        +\frac{21f''f^{(5)}}{(y')^8}
        +\frac{35f'''f''''}{(y')^8}
        -\frac{210(f'')^2f''''}{(y')^9}
                \continue
& &     -\frac{280f''(f''')^2}{(y')^9}
        +\frac{1260(f'')^3f'''}{(y')^{10}}
        -\frac{945(f'')^5}{(y')^{11}}
\,.
\label{F_5rde}
\eea

Perturbative formulas for 5th and 6th order conjugacy coefficients,
continuation of \refeq{h(5)}:
{\small
\bea
b_5 &=&
2 (7 + 3\Lambda + 2\Lambda^2 ) f_2^4
    + \Lambda  \left( -21 + \Lambda  
    + 6{{\Lambda }^2} + 11{{\Lambda }^3} + 3{{\Lambda }^4} \right) 
      {f_2^2}f_3
	\ceq
    + 2\Lambda^2{{\left( \Lambda -1 \right) }^2} 
      \left( 3 + 5\Lambda  + 4{{\Lambda }^2} + 2{{\Lambda }^3} \right) 
      f_2f_4 
	+ 3 \Lambda^2 (\Lambda - 1) (\Lambda^3 - 1) f_3^2 
	\ceq
	+ \Lambda^3 (\Lambda - 1) (\Lambda^2 - 1) (\Lambda^3 - 1) f_5
\continue
b_6 &=&
2\left(21+14\Lambda+14{\Lambda^2}+8{\Lambda^3}+3{\Lambda^4}\right)
{f_2^5}\ceq
+\Lambda\left(-84-21\Lambda-6{\Lambda^2}+25{\Lambda^3}+
44{\Lambda^4}+27{\Lambda^5}+14{\Lambda^6}+{\Lambda^7}\right){f_2^3}
f_3\ceq
+2{\Lambda^2}\left(14+\Lambda-7{\Lambda^2}-9{\Lambda^3}-
11{\Lambda^4}+{\Lambda^5}+{\Lambda^6}+7{\Lambda^7}+3{\Lambda^8}\right)
{f_2^2}f_4\ceq
+{{\left(-1+\Lambda\right)}^2}{\Lambda^2}f_2
\left[\left(28+43\Lambda+50{\Lambda^2}+43{\Lambda^3}+22{\Lambda^4}+
6{\Lambda^5}\right){f_3^2}\right.\ceq\left.
+\Lambda\left(-7-12\Lambda-10{\Lambda^2}-3{\Lambda^3}+7{\Lambda^4}+
10{\Lambda^5}+10{\Lambda^6}+5{\Lambda^7}\right)f_5\right]\ceq
+{{\left(-1+\Lambda\right)}^3}{\Lambda^3}
\left(1+\Lambda+{\Lambda^2}+{\Lambda^3}\right)\ceq
\times\left[\left(7+7\Lambda+4{\Lambda^2}\right)f_3f_4
+\Lambda\left(-1-\Lambda+{\Lambda^3}+{\Lambda^4}\right)f_6\right]
\,.
\label{h(5)a}
\eea
} 
Continuation of the derivative evaluation
\refeq{der(3)} - 
the 4th and 5th order derivatives:
{\small
\bea
\left.\frac{1}{5!}\frac{\partial^4}{\partial y^4}
\frac{1}{y'(x)}\right|_{x=0}
 &=&
{\frac{{{\Lambda }^r} \left( 1 + {{\Lambda }^r} \right) }{
       {{\left( {{\Lambda }^r} -1 \right) }^5}}}  
      \left\{ 14 {{\left( 1 + {{\Lambda }^r} \right) }^2} {h_2^4} 
        \right.
\label{der(5)a} 
	\\ & &
	- 3 \left( 7 + 10 {{\Lambda }^r} + 7 {{\Lambda }^{2 r}} \right)  
         {h_2^2} h_3
        + 3 \left( 1 + {{\Lambda }^r} + {{\Lambda }^{2 r}} \right)  {h_3^2}
        \ceq
	\left.
        + 2 \left( 3 + 2 {{\Lambda }^r} + 3 {{\Lambda }^{2 r}} \right)  
	h_2  h_4
	- \left( 1 + {{\Lambda }^{2 r}} \right)  h_5
	\right\}
        \continue
\left.\frac{1}{6!}\frac{\partial^5}{\partial y^5}
\frac{1}{y'(x)}\right|_{x=0}
 &=&
-\frac{\Lambda^r }{ {\left( \Lambda^r -1 \right)}^6 }
 \left\{
          42 {{\left( 1 + {{\Lambda }^r} \right) }^4} 
         {h_2^5} 
        \right.
        \ceq
	- 28 {{\left( 1 + {{\Lambda }^r} \right) }^2} 
         \left( 3 + 4 {{\Lambda }^r} + 3 {{\Lambda }^{2 r}} \right)  
	{h_2^3}  h_3
        \ceq
	+ 14 {{\left( 1 + {{\Lambda }^r} \right) }^2} 
         \left( 2 + {{\Lambda }^r} + 2 {{\Lambda }^{2 r}} \right)  
	{h_2^2} h_4
        \ceq
	-  \left( 7 + 14 {{\Lambda }^r} + 18 {{\Lambda }^{2 r}} + 
           14 {{\Lambda }^{3 r}} + 7 {{\Lambda }^{4 r}} \right)  h_3 
         h_4 
        \ceq
	+ 7 \left( 4 + 11 {{\Lambda }^r} + 15 {{\Lambda }^{2 r}} + 
              11 {{\Lambda }^{3 r}} + 4 {{\Lambda }^{4 r}} \right) h_2 {h_3^2}
        \ceq
             - \left( 7 + 12 {{\Lambda }^r} + 12 {{\Lambda }^{2 r}} + 
              12 {{\Lambda }^{3 r}} + 7 {{\Lambda }^{4 r}} \right) h_2 h_5
        \ceq
	\left.
	 + \left( 1 + {{\Lambda }^r} + {{\Lambda }^{2 r}} + 
           {{\Lambda }^{3 r}} + {{\Lambda }^{4 r}} \right)  h_6 
\right\}
\nnu
\,.
\eea
} 
The  two-point fourth order correction (continuation of
calculations of \refsect{s-Two-p-int}):
\begin{eqnarray}
& &
\left.
\frac{\partial^2}{\partial y_j^2}
\frac{\partial^2}{\partial y_k^2}
\frac{1}{|(\partial f^{j}(x_a)/\partial x_a)
( \partial f^{k}(x_b)/\partial x_b)-1|}\right|_{\rm cycle}
	\continue
&=&\frac{\Lambda}{(\Lambda-1)^5}\left\{(1+\Lambda)\left[\Lambda_k^2
\left(60h_a^{(2)^4}-72h_a^{(2)^2}h_a^{(3)}+12h_a^{(2)}h_a^{(4)}+9h_a^{(3)^2}
-h_a^{(5)}\right)\right.\right.
	\continue
&&\left.+
\Lambda_j^2\left(60h_b^{(2)^4}-72h_b^{(2)^2}h_b^{(3)}+12h_b^{(2)}h_b^{(4)}
+9h_b^{(3)^2}-h_b^{(5)}\right)\right]\label{2p4}\\
&+&(3+10\Lambda+3\Lambda^2)\left[\Lambda_kh_b^{(2)}\left(
12h_a^{(2)^3}-9h_a^{(2)}h_a^{(3)}+h_a^{(4)}\right)\right.
	\continue
&&\left.+\Lambda_jh_a^{(2)}\left(
12h_b^{(2)^3}-9h_b^{(2)}h_b^{(3)}+h_b^{(4)}\right)\right]
	\continue
&+&\left.(1+\Lambda)(1+10\Lambda+\Lambda^2)\left(3h_a^{(2)^2}-h_a^{(3)}\right)
\left(3h_b^{(2)^2}-h_b^{(3)}\right)\right\}
\,,
\nonumber
\end{eqnarray}
and the two-point fifth order correction is:
{\small
\begin{eqnarray}
& &
\left.
	\frac{\partial^3}{\partial y_j^3}
	\frac{\partial^2}{\partial y_k^2}
	\frac{1}{|(\partial f^{j}(x_a)/\partial x_a)
	(\partial f^{k}(x_b)/\partial x_b)-1|}\right|_{\rm cycle}
	\continue
	&=&\frac{\Lambda}{(\Lambda-1)^6}\left\{
	(1+\Lambda)\Lambda_k^3\left(-360h_a^{(2)^5}+600h_a^{(2)^3}h_a^{(3)}
	-180h_a^{(2)}h_a^{(3)^2}
			\right. \right.   \ceq \left. \left. 
	-120h_a^{(2)^2}h_a^{(4)}+30h_a^{(3)}h_a^{(4)}
	+15h_a^{(2)}h_a^{(5)}-h_a^{(6)}\right)\right. \continue
	&+&6(1+3\Lambda+\Lambda^2)\Lambda_k^2h_b^{(2)}
	\left(-60h_a^{(2)^4}+72h_a^{(2)^2}h_a^{(3)}-9h_a^{(3)^2}
	-12h_a^{(2)}h_a^{(4)}+h_a^{(5)}\right)   \continue
	&+&\left[15(1+7\Lambda+7\Lambda^2+\Lambda^3)h_b^{(2)^2}
	-2(2+13\Lambda+13\Lambda^2+2\Lambda^3)h_b^{(3)}\right]
			\ceq
\Lambda_k \left(-12h_a^{(2)^3}+9h_a^{(2)}h_a^{(3)}-h_a^{(4)}\right) \continue
&+&\left[15(1+18\Lambda+42\Lambda^2+18\Lambda^3+\Lambda^4)h_b^{(2)^3}
			\right. \label{2p5th} \\& & \left. 
-2(5+82\Lambda+186\Lambda^2+82\Lambda^3+5\Lambda^4)h_b^{(2)}h_b^{(3)}
			\right.  \ceq \left. 
+(1+14\Lambda+30\Lambda^2+14\Lambda^3+\Lambda^4)h_b^{(4)}\right]
\left(-3h_a^{(2)^2}+h_a^{(3)}\right) \continue
&+&3\left[-30(3+17\Lambda+17\Lambda^2+3\Lambda^3)h_b^{(2)^4}
+5(19+101\Lambda+101\Lambda^2+19\Lambda^3)h_b^{(2)^2}h_b^{(3)}
                        \right.  \ceq \left.
-10(1+5\Lambda+5\Lambda^2+\Lambda^3)h_b^{(3)^2}\right. \continue
&&\left.-2(7+33\Lambda+33\Lambda^2+7\Lambda^3)h_b^{(2)}h_b^{(4)}
+(1+4\Lambda+4\Lambda^2+\Lambda^3)h_b^{(5)}\right]\Lambda_jh_a^{(2)} \continue
&+&\left[-630(1+2\Lambda+\Lambda^2)h_b^{(2)^5}
+30(31+58\Lambda+31\Lambda^2)h_b^{(2)^3}h_b^{(3)}
			\right.  \ceq \left. 
-60(4+7\Lambda+4\Lambda^2)h_b^{(2)}h_b^{(3)^2}
-15(11+18\Lambda+11\Lambda^2)h_b^{(2)^2}h_b^{(4)}\right. \continue
&&\left.\left.+2(17+26\Lambda+17\Lambda^2)h_b^{(3)}h_b^{(4)}
+6(3+4\Lambda+3\Lambda^2)h_b^{(2)}h_b^{(5)}
-(1+\Lambda+\Lambda^2)h_b^{(6)}\right]\Lambda_j^2\right\}
\,.
\nonumber
\end{eqnarray}
} 
The repetition number dependent prefactors in
\refeq{der(3)} are turned into power series in $\Lambda^{rk}$
using identities
\beq
        {1+x \over (1-x)^3} = \sum_{k=0}^{\infty} (k+1)^2 x^k
\ee{sum0}

\beq
{\frac{ 5 + 8 x  + 5 x^2 }
     {{{\left( 1  -x \right)
         }^4}}}
=
\sum_{k\geq 0} 
	\left( 1 + k \right)  
  \left( 5 + 6 k + 3 {k^2} \right) x^k
\ee{sumB}

\beq
\frac{1 + x  + x^2}
     {\left( 1 -x \right)^4}
=
\frac{1}{2}\sum_{k\geq 0} 
	\left( 1 + k \right)   \left( 2 + 2 k + {k^2} \right) x^k
\ee{sumC}

\end{document}